\documentclass[letterpaper, 10 pt, conference]{ieeeconf}  

\IEEEoverridecommandlockouts           

\usepackage{cite}
\usepackage{amsmath,amssymb,amsfonts}
\usepackage{graphicx}
\usepackage{textcomp}
\usepackage{xcolor}
\def\BibTeX{{\rm B\kern-.05em{\sc i\kern-.025em b}\kern-.08em
    T\kern-.1667em\lower.7ex\hbox{E}\kern-.125emX}}
\usepackage[utf8]{inputenc}
\usepackage{amsmath,graphicx}
\usepackage{mathrsfs}
\usepackage{xcolor}
\usepackage{mathtools}
\usepackage{enumerate}
\usepackage{booktabs}
\usepackage{romannum}
\usepackage{float}
\usepackage{algorithm}
\usepackage{algpseudocode}

\usepackage{setspace}
\renewcommand{\baselinestretch}{0.997}

\usepackage{amsthm}
\newtheorem{theorem}{Theorem}
\newtheorem{lm}{Lemma}

\newtheorem{conj}{Conjecture}

\newtheorem{rem}{Remark}

\newcommand{\edit}[1]{\textcolor{black}{#1}}

\usepackage[normalem]{ulem}

\newcommand{\x}[1]{\mathbf{x}_{#1}}
\newcommand{\vvec}[1]{\mathbf{v}_{#1}}
\newcommand{\z}[1]{\mathbf{z}_{#1}}

\makeatletter
\renewcommand*\env@matrix[1][\arraystretch]{%
  \edef\arraystretch{#1}%
  \hskip -\arraycolsep
  \let\@ifnextchar\new@ifnextchar
  \array{*\c@MaxMatrixCols c}}
\makeatother

\newcommand{\RR}{{\mathbb{R}}}

\DeclareUnicodeCharacter{2212}{-}
\title{\LARGE \bf
Defending a Static Target Point with a Slow Defender
}

\author{Goutam Das$^{1}$, Michael Dorothy$^{2}$, Zachary I. Bell$^{3}$, and Daigo Shishika$^{4}$
\thanks{We gratefully acknowledge the support of  ARL grant ARL DCIST CRA
W911NF-17-2-0181 and AFRL grant FA8651-23-1-0012.
The views expressed in this paper are those of the authors and do not reflect the official policy or position of the United States Government, Department of Defense, or its components.}
\thanks{$^{1}$Goutam Das, PhD student, Electrical and Electronics Engineering, George Mason University, 4400 University Dr, Fairfax, VA 22030, USA
        {\tt\small gdas@gmu.edu}}%
\thanks{$^{2}$Michael Dorothy. Army Research Directorate, DEVCOM Army Research Laboratory, APG, MD.
        {\tt\small michael.r.dorothy.civ@army.mil}}%
\thanks{$^{3}$Zachary I. Bell is with the Munitions Directorate, Air Force Research Laboratory, Eglin AFB, FL 32542, USA {\tt\small zachary.bell.10@us.af.mil.}}%
\thanks{$^{4}$Daigo Shishika, Assistant Professor, Department of Mechanical Engineering, George Mason University,
        4400 University Dr, Fairfax, VA 22030, USA
        {\tt\small dshishik@gmu.edu}}%
}

\begin{document}
\maketitle
\thispagestyle{empty}
\pagestyle{empty}


\begin{abstract}
This paper studies a target-defense game played between a slow defender and a fast attacker. 
The attacker wins the game if it reaches the target while avoiding the defender's capture disk.
The defender wins the game by preventing the attacker from reaching the target, which includes reaching the target and containing it in the capture disk.
Depending on the initial condition, the attacker must circumnavigate the defender's capture disk, resulting in a constrained trajectory. This condition produces three phases of the game, which we analyze to solve for the game of kind.
We provide the barrier surface that divides the state space into attacker-win and defender-win regions, and present the corresponding strategies that guarantee win for each region.
Numerical experiments demonstrate the theoretical results as well as \edit{the efficacy of the proposed strategies.}
\end{abstract}
\section{Introduction}
%
Pursuit-evasion games (PEGs) have \edit{emerged as a significant area of} research \edit{within the domains} of control theory, artificial intelligence, and robotics, among others. \edit{This interest is driven by their wide-ranging} applications in both civilian and military \edit{contexts}. These problems \edit{can be further categorized into subfields such} as reach-avoid games, target-defense games, and perimeter-defense games, \edit{to name a few} \cite{von2021turret,weintraub2020introduction}.

In reach-avoid games (RAGs), the players are divided into pursuers and evaders, each with the objective of reaching a 
target before the opposing player while avoiding or maintaining specific constraints \cite{pachter2020capture}.

Target-defense games (TDGs) are a type of PEGs that were first introduced by \edit{Isaacs} in his seminal work \cite{Issacs1965}. In TDGs, attacker(s) seek to reach a high-value asset (HVA) or target, and defender(s) seek to protect the target from the incoming attacker \cite{shishika2021partial,pachter2019toward}.
Targets can be a simple static point in 2D space,  \edit{a circle\cite{VonMoll2020BD}, or a moving line \cite{das2022guarding}.}
\edit{A static target} with 
arbitrary convex shape \edit{has} been explored in \cite{shishika2019perimeter}.
A generalized version of TDGs where the target is in $n$-dimensional space is solved in \cite{lee2021guarding}.
 Depending on the number of players, the games are classified \edit{into} single defender many attackers, two defenders single attacker, or many defender many attacker games \cite{pourghorban2023target,Garcia2021Multiple}.
 
One key assumption in most of these problems is that the defenders are, in general, at least as fast as the attackers.  
A common approach to handle slower-defender target-defense games (SD-TDGs) is the \emph{encirclement} of the faster attacker \cite{garcia2021cooperative_containment, ramana2015cooperative, vechalapu2020trapping, fang2020cooperative}, which requires a significant number of defenders depending on the speed ratio parameter.
SD-TDG with only one defender is \edit{a}
 much less explored area in the literature of TDGs.
Cooperative TDG with \edit{a} single defender is explored in \cite{garcia2021cooperative_targetprotection} where the target cooperates with the defender. 
A special case of the SD-TDG was studied in \cite{pachter2022strategies}, where the defender \edit{employs} an open-loop strategy to go straight to the target, and the attacker is on the capture radius which circumnavigates around the defender

\edit{The work most closely related to this paper is found in \cite{fu2022defending}, where the defender aims to perpetually prevent the attacker from reaching the target region.}
The strategies used in the paper are derived from \cite{fu2021optimal} which considers a target-defense game with multiple defenders with a similar approach.





We formulate the problem \edit{of defending a point target as a reach-avoid game, where the target is defended if the defender reaches it first, instead of having to perpetually circle around and block.}
This modification led to some key differences in the derived strategies as well as the resultant winning regions (i.e., barrier surface).
The first key difference of our work compared to \cite{fu2022defending,fu2021optimal} is that our defender strategy never moves away from the target in an effort to capture the attacker, whereas the ones derived in \cite{fu2022defending,fu2021optimal} led to such behavior even for a point target scenario.
Secondly, our analysis accommodates the entire state space compared to~\cite{fu2022defending} that worked only for defender that is close enough to the target.  Thirdly, the barrier surface is always closed and it always provides a compact defender-winning region, compared to a barrier segment which requires multiple defender to close the barrier around the target in \cite{fu2022defending}.
A comparative study is provided to demonstrate these differences at the end.

The paper is organized into the following sections: Section~II formulates the SD-TDG; Section~III introduces dominance regions to identify trivial winning regions; Section~IV 
derives optimal strategies \edit{for} when the defender is able to ``block'' the attacker, i.e., when the attacker must take a constrained trajectory avoiding defender's capture circle;
Section~V discusses the construction of the closed barrier surface and provides security strategies inside the winning regions; and Section~VI compares our results with other existing literature.

\section{Problem Formulation}
\begin{figure}\label{fig: intro}
    \centering
    \includegraphics[width = 0.40 \textwidth]{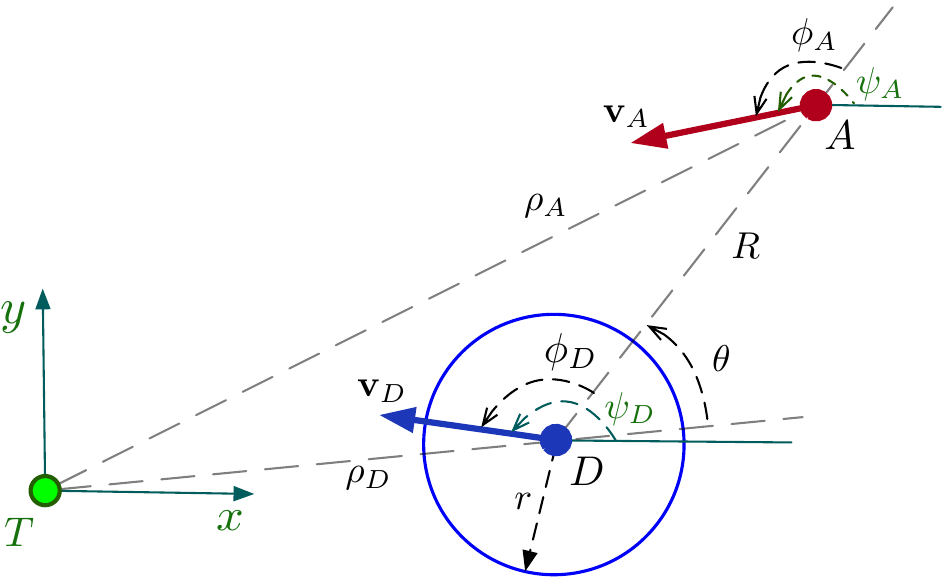}
    \caption{Illustration of the target-defense game with a slower-defender.}
    \label{fig: prob_formulation}
\end{figure}
This section formulates the slower\edit{-}defender target\edit{-}defense game (SD-TDG) on a plane. 
\edit{The positions of the agents are denoted as $\x{i} = [x_i, y_i]$, where $i \in \{T, D, A\}$} represents the target, the defender ($D$), and the attacker ($A$), respectively. 
\edit{Without loss of generality, we consider the static target ($T$) to be at the origin, i.e., $\x{T} = [0,0]^\top$.} 
\edit{The agents} have first-order dynamics as follows:
\begin{align}
\begin{aligned}
\dot{\z{}} &= f(\z{}) =
    \begin{bmatrix}
    \dot{x}_D  \\
    \dot{y}_D  \\
    \dot{x}_A  \\ 
    \dot{y}_A  
    \end{bmatrix} =     
    \begin{bmatrix}
    \nu\cos{\psi_D} \\ 
    \nu\sin{\psi_D} \\
    \cos{\psi_A} \\
    \sin{\psi_A} 
    \end{bmatrix},
\end{aligned}
\end{align}
where $\z{} = [x_D,y_D, x_A, y_A]^\top$ \edit{represents} the \edit{stacked} state of the game. $\psi_D, \psi_A \in [-\pi,\pi)$ are the heading angles (control action) of the defender and the attacker, \edit{respectively}.
\edit{The corresponding unit vectors of the defender and attacker headings are denoted by $\mathbf{v}_D=\left[\cos \psi_D, \sin \psi_D\right]^{\top}$, and $\mathbf{v}_A=\left[\cos \psi_A, \sin \psi_A\right]^{\top}$, respectively. The speed ratio parameter, $\nu \in (0,1)$, indicates that the defender is slower than the attacker.}
Additionally, the defender has a \emph{capture radius}
$r>0$. We define the distance between the players at time $t$ as $R(t) \triangleq \|\x{D}(t)-\x{A}(t)\|$,
but the time argument will be omitted for conciseness. 
The \edit{instantaneous} capture condition is $R < r$, which the attacker must avoid. \edit{The scenario is modeled as a full-state information game, i.e., all players are aware of each other's instantaneous states at any given time. However, a player's instantaneous control action (the heading angle) is not known to the other player.}

We consider a \emph{game of kind} where the attacker wins the game if it reaches the target while avoiding defender's capture disk. \edit{Alternatively,} the defender wins if it can prevent the attacker from reaching the target by containing the target within its capture disk. \edit{For simplicity, we refer to this case as `defender reaches the target' before the attacker.}

Following the notation used in
\cite{fu2021optimal}, we use $\rho_D$ and $\rho_A$ to denote the distances of the defender and the attacker from the target, at any given time $t$: i.e., $\rho_D(t) \triangleq \|\x{D}(t)\|$ and $\rho_A(t) \triangleq \|\x{A}(t)\|$. 
The game ends when either  $\rho_D(t_f) = r$ (defender win), or  $\rho_A(t_f) = 0$ (attacker win), whichever \edit{occurs} first, where $t_f$ is the terminal time. The terminal surfaces $\mathcal{S}_D$ and  $\mathcal{S}_A$ corresponding to defender win and attacker win respectively are then given as follows:
\begin{align}
    \mathcal{S}_D = \{\z{} \mid \rho_{D} = r, \quad  \rho_{A} > 0\}, \\
    \mathcal{S}_A = \{\z{} \mid \rho_{A} = 0, \quad \rho_{D} > r\}.
\end{align}
We use \(\mathcal{R}_D\) and \(\mathcal{R}_A\) to denote the winning regions of the defender and the attacker, \edit{respectively.} From \(\mathcal{R}_D\) (resp. \(\mathcal{R}_A\)), the defender (resp. attacker) has a strategy to guarantee its win. The surface that divides these two regions is called the \emph{barrier surface}, denoted as \(\mathcal{B}\). \edit{It is characterized by the set of initial conditions from which the game terminates at \(t_f\) with \(\rho_D = r\) and \(\rho_A = 0\) simultaneously.} 
The main objective of this paper is to identify the barrier surface \(\mathcal{B}\) and the corresponding winning strategies for the players.

\section{Dominance Regions}\label{sec: dominance}
We first look into the \emph{dominance regions} of the players. 
A dominance region for a player contains all the points it can reach \emph{first} in a straight-line regardless of the opponent's strategy.
This \edit{analysis} \edit{yields a} region in the state space that results in \edit{a} trivial answer to the game of kind. 

\subsection{Defender's Dominance Region}
\begin{figure}
    \centering
    \includegraphics[width = 0.35\textwidth]{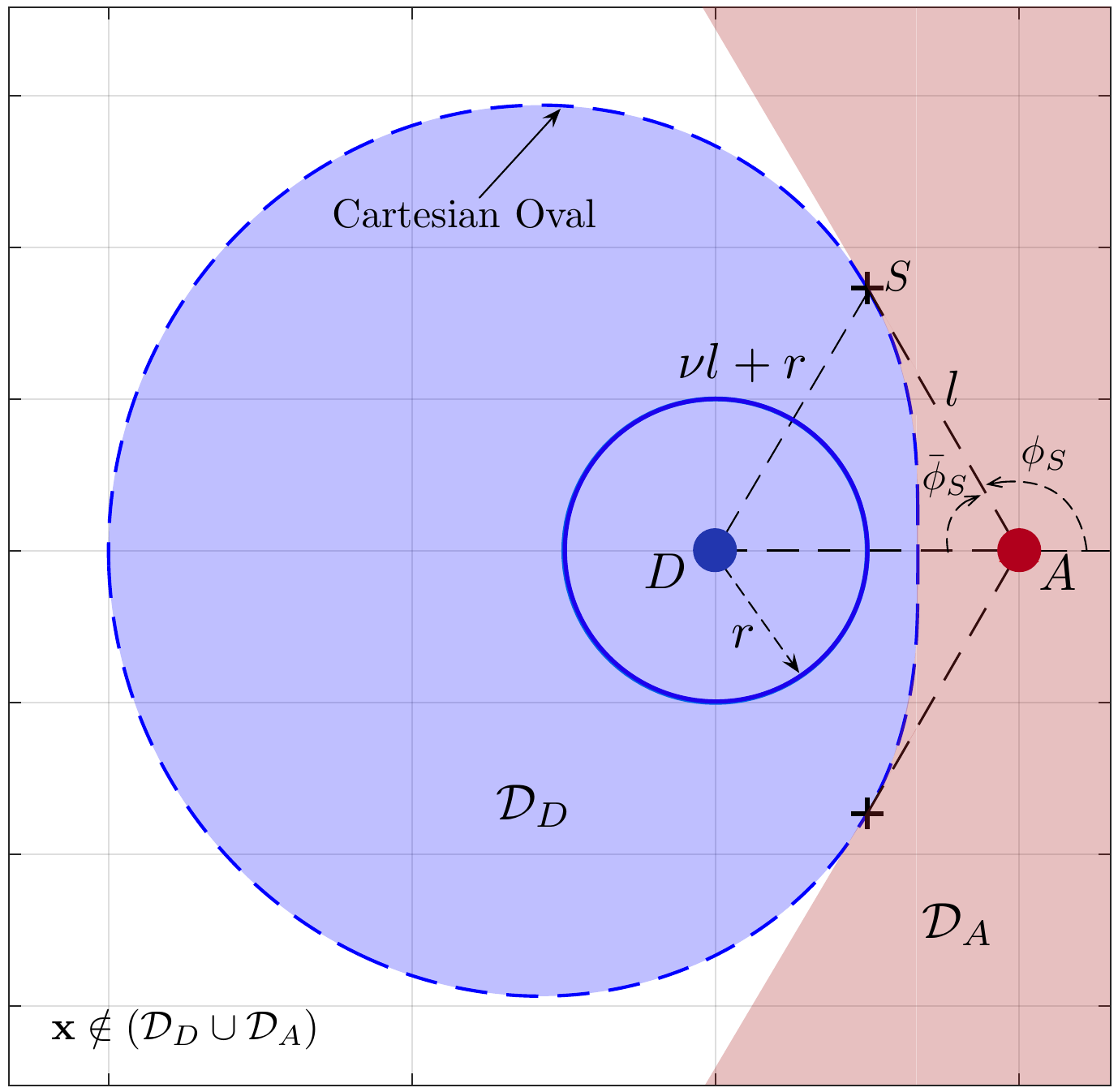}
    \caption{Defender's dominance region, $\mathcal{D}_D$ is shown as a Cartesian Oval (CO) and the attacker's dominance region, $\mathcal{D}_A$ is shown as a shaded region outside the CO. 
    } 
    \label{fig:CO}
\end{figure}

Since the defender has an extended ``reach'' of $r$, its dominance region in free space is defined as follows:
\begin{align}\label{eq: D_D}
    \mathcal{D}_D \triangleq \{\x{} \in \RR^2 \mid \|\x{} - \x{D}\|-r \leq \nu \|\x{} - \x{A}\| \}.
\end{align}
The boundary of \eqref{eq: D_D} is \edit{referred to} as the \emph{Cartesian Oval} (CO) shown in Figure~\ref{fig:CO}, for which the following result is known \cite{garcia2021cooperative_containment}.
\begin{lm}[Lemma~2 in \cite{garcia2021cooperative_containment}]\label{lemma: CO}
    The CO for the slower defender and the faster attacker is described by the following equations:
    \begin{align}
        x &= x_A + l\cos{(\bar{\phi}+\lambda)}, \\
        y &= y_A + l\sin{(\bar{\phi}+\lambda)},
    \end{align}
    where $l$ is the distance of a point on CO from the attacker:
    \begin{align}
        l &= \dfrac{\eta \pm \sqrt{\eta^2 - (1-\nu^2)(R^2-r^2)}}{1-\nu^2},
    \end{align}
    for $\bar{\phi} \in [-\bar{\phi}_S, \bar{\phi}_S]$, 
    \begin{align}\label{eq: phi_S}
        \bar{\phi}_S \triangleq \pi - \phi_S = \arccos{\left(\frac{\sqrt{(1-\nu^2)(R^2-r^2)}-\nu r}{R}\right)},
    \end{align}
    and 
    $\lambda$ is the line-of-sight (LOS) angle,
\begin{align}
\lambda &= \arctan{\left(\frac{y_D-y_A}{x_D-x_A}\right)}, \label{eq: lambda} \text{ and }
\\
\eta &= \nu r + R\cos{\bar{\phi}}.
\end{align}
\end{lm}

\begin{rem}
    The defender has a strategy to reach any point inside the CO before the attacker does, thus if the target is inside this region, the defender can win the game.
\end{rem}

\begin{rem}
\label{rem:CO_for_attacker}
The CO is merely a geometric construction which \edit{outlines} all points $\x{}$ that
satisfy the condition: $\frac{\|\x{}-\x{D}\|-r}{\|\x{}-\x{A}\|}=\nu$.
It does not consider the capture of attacker if the trajectories goes through the defenders \edit{capture disk}. 
\edit{Consequently, the CO offers no }guarantee for the attacker to reach a target 
\edit{within} or beyond the CO (\edit{e.g.,}  the white region in Figure~\ref{fig:CO}). 
\end{rem}

\subsection{Attacker's Dominance Region} 
Due to Remark~\ref{rem:CO_for_attacker}, identifying the exact dominance region for the attacker is more subtle.
We find a subset of the dominance region as follows:
\begin{equation}\label{eq: D_A}
    \mathcal{D}_A=\{\x{} \;|\; \mu\x{A}+(1-\mu)\x{} \notin \mathcal{D}_D,\forall \mu\in[0,1]\}.
\end{equation}
In words, $\mathcal{D}_A$ contains a set of points that the attacker can reach without passing through the CO.
This region is indicated as the red shaded area in Figure~\ref{fig:CO}.

\begin{lm}\label{lem: D_A}
If the target is in $\mathcal{D}_A$, then the attacker can win the game.    
\end{lm}
\begin{proof}
\edit{According to} definition \eqref{eq: D_D}, the attacker \edit{can} reach any point outside CO before the defender does, \edit{provided that there is no interception within the defender's \edit{capture disk}.}
The \edit{possibility of capture}
only arises if the attacker's trajectory passes through $\mathcal{D}_D$, and therefore, any point in $\mathcal{D}_A$ is safely reachable by the attacker.
\end{proof}

In terms of 
the relative location of the target, we have now carved out two regions $\mathcal{D}_D$ and $\mathcal{D}_A$ from our analysis since they provide
immediate solution to the game of kind.
When the target is outside of $\mathcal{D}_D\cup\mathcal{D}_A$ (i.e., the 
\edit{white} region in Figure~\ref{fig:CO}), we know 
\edit{from Remark~\ref{rem:CO_for_attacker},} that the defender has a strategy to position itself so that the attacker must \edit{adopt}
a non-straight path in order to reach the target. 

We call such a situation as ``blocking'' where the defender places itself between the target and the attacker, 
\edit{compelling} the attacker 
\edit{to} take a detour and \edit{consequently, a} longer time to reach the target. 
\edit{This poses a tactical dilemma: \emph{How should the defender optimally allocate its velocity components between blocking the attacker and moving towards the target?}
}

\section{The Blocking Game}\label{sec: IV blocking game}
In this section we will discuss the case when the defender can block the attacker from reaching the target: $\x{T} \notin \mathcal{D}_D \cup \mathcal{D}_A$. 
To derive the optimal strategy and subsequently the reachable region of the players, we first break the trajectories into three possible phases.
\begin{itemize}
    \item \textbf{Phase-I:}  
    $R > r$ and $\x{T}\notin\mathcal{D}_D\cup \mathcal{D}_A$, i.e., the attacker is outside the \edit{capture disk} of the defender. There is no trivial straight-line trajectory for either player to win.
      
    \item \textbf{Phase-II:} $R = r$ and $\x{T}\notin\mathcal{D}_D\cup \mathcal{D}_A$, i.e., the attacker is on the \edit{capture disk} of the defender. The attackers objective is twofold: (i) avoid capture, and (ii) circumnavigate the defender until $\x{T} \in \mathcal{D}_A$.
    
    \item \textbf{Phase-III:} $R \geq r$ and $\x{T}\in\mathcal{D}_D\cup \mathcal{D}_A$: 
    final stage of the game where either the defender or the attacker has a 
    \edit{straight-line path} to win.

\end{itemize}

We will build our strategies in reverse chronological order from the terminal surface. However, since the analysis of Phase-III is trivial, 
\edit{our primary focus will be on}
Phase-II. For this reason, we start by introducing the dynamics in Phase-II first.

\subsection{Dynamics of Phase-II}
\begin{figure}
    \centering
    \includegraphics[width = 0.3\textwidth]{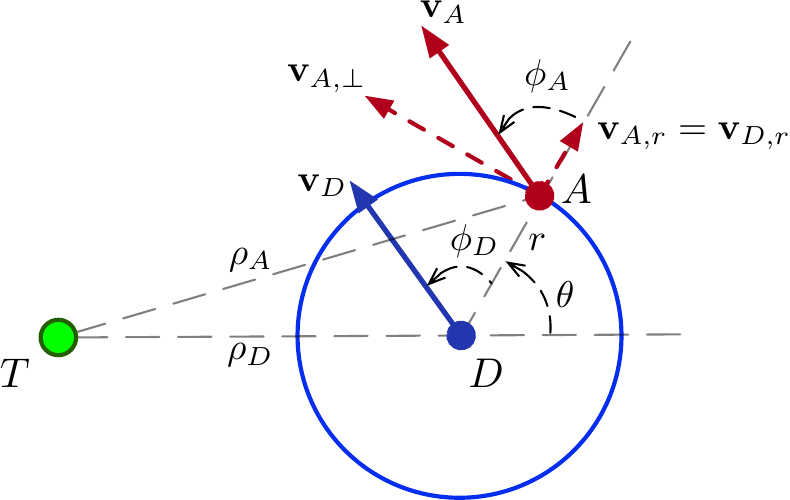}
    \caption{The defender and the attacker in $(\rho_D,\theta)$ space in Phase-II.
    } 
    \label{fig: rhoD-theta}
\end{figure}

Due to the rotational symmetry around the target and the distance constraint $R(t)=r$, we can reduce the four-dimensional state space to just two dimension.
We \edit{choose}
$\rho_D$ and $\theta$ to be the states, where $\theta = \pi- \angle TDA$ \edit{represents}
the angular position of the attacker with respect to the $TD$-axis.
For convenience, the control variables are
\edit{denoted as}
$\phi_D$ and $\phi_A$, as shown in Figure~\ref{fig: rhoD-theta}.

The kinematics of the players in this stage can be described by the following equations:
\begin{align}
    \dot{\rho}_D &= \nu\cos{(\theta+\phi_D)}, \label{eq: kinematics rhoDdot}\\
    \dot{\theta} &= -\frac{\nu}{\rho_D}\sin{(\theta+\phi_D)}+\frac{1}{r}\left( \sin{\phi_A}-\nu\sin{\phi_D}\right).\label{eq: kinematics thetaDot}
\end{align}

\edit{To maintain constant distance between the players} (i.e., $\dot{R} =0$), we impose the following constraint on the control of the attacker:
\begin{align}\label{eq: p2_heading_constraint}
    \nu\cos{\phi_D} = \cos{\phi_A}.
\end{align}
Note that the same assumption is made in other related works \cite{fu2021optimal, hagedorn1976differential}.
Although the attacker needs access to the defender's instantaneous heading \edit{and speed}
to achieve this constraint, it is reasonable in the context of Nash equilibrium; i.e., the attacker can compute the equilibrium heading of the defender.
Furthermore, since the attacker is faster than the defender, any attempt to capture the attacker can be safely avoided if the attacker uses some buffer; $r_\text{safe} = r +\epsilon$, where $0<\epsilon<<r$. For simplicity, we proceed with $r_\text{safe} = r$.

\subsection{Phase-III / Termination of Phase-II}
We describe conditions on $(\rho_D,\theta)$ that immediately lead to the winning of one player. This effectively extends the terminal surfaces $\mathcal{S}_D$ and $\mathcal{S}_A$ of the overall game to the termination of Phase-II.

\subsubsection{Defender Winning Terminal Condition}
\begin{figure}
    \centering
    \includegraphics[width = 0.35\textwidth]{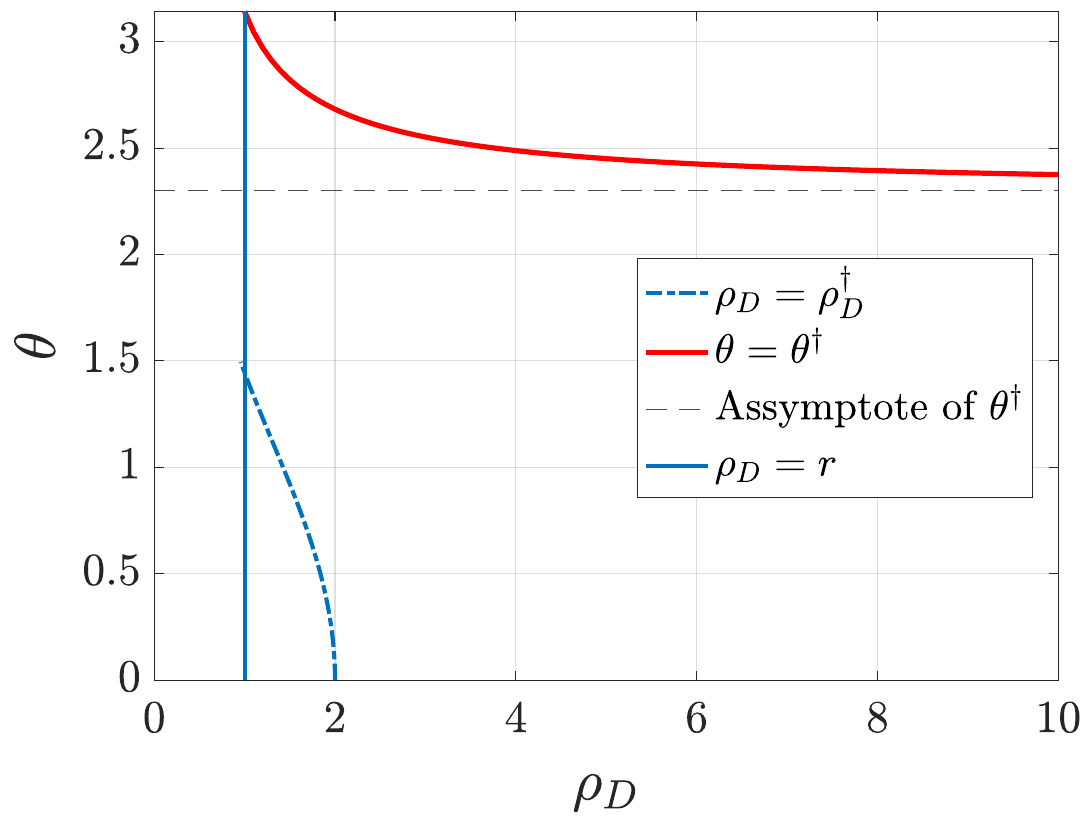}
    \caption{Terminal surfaces of the Phase-II indicating the end of game trivially by either defender or attacker win.} 
    \label{fig: terminal_surface}
\end{figure}

The first obvious condition equivalent to $\mathcal{S}_D$ is $\rho_D = r$. This is depicted as the solid blue line in Figure~\ref{fig: terminal_surface}.

The other condition for the defender to win is when it has advantage in angular speed.  
Define $\rho_{D}^\dagger(\theta)$ as the critical distance of the defender from the target such that the defender can reach the target while blocking the attacker due to angular speed advantage.
\begin{figure}
    \centering
    \includegraphics[width = 0.37\textwidth]{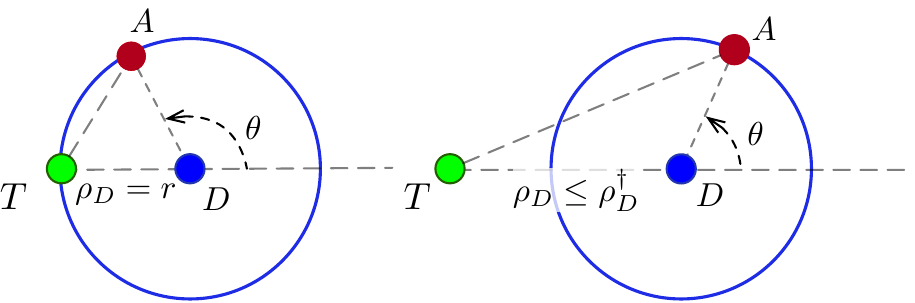}
    \caption{Defender-win terminal conditions: defender reaches the target (left), 
 and attacker is blocked while the defender has angular advantage (right).}
    \label{fig: D-win conditions}
\end{figure}

\begin{theorem}[Defender Win Condition]\label{theorem: 1} The defender has a strategy to win the game if $\rho_D \leq \rho_D^\dagger(\theta)$, where 
\begin{align}\label{eq: rhoD_perp}
        \edit{\rho_D^{\dagger}(\theta) \triangleq \frac{\nu r\left(\nu \cos \theta+\sqrt{1-\nu^2 \sin ^2 \theta}\right)}{1-\nu^2}} .
\end{align}
\end{theorem}

\begin{proof}
    \edit{If the target is not inside the attackers dominance region, i.e.,  $\x{T} \notin \mathcal{D}_A$, the attacker must go around the defender to reach it. In Phase-III, \edit{if $\rho_D > r$} for the defender win case, the attacker will be on the \edit{capture disk} of the defender}
    and seek to go around, it must increase $|\theta|$. 
    However, the defender can prevent $|\theta|$ from increasing, while reducing $\rho_D$, if it has angular speed advantage around the target.
    This is true if the distances satisfy the following condition:
    \begin{align} \label{eq: D_perp block}
        \frac{\nu}{\rho_D}\geq \frac{1}{\rho_A}.
    \end{align}
    Since the attacker is on the disk, the geometry in Figure~\ref{fig: D-win conditions}-(right) provides
    \begin{align}\label{eq: rho_A,perp}
            \rho_A = \sqrt{(\rho_D + r\cos{\theta})^2 + r^2\sin^2{\theta}} \geq 0.
    \end{align}
    From the critical value in \eqref{eq: D_perp block} and~\eqref{eq: rho_A,perp}, we get \eqref{eq: rhoD_perp}, which gives the critical defender distance, $\rho_D = \rho_{D}^\dagger(\theta)$, for a given attacker position on the \edit{capture disk}. 
\end{proof}

\subsubsection{Attacker Winning Terminal Condition}
\begin{figure}
    \centering
    \includegraphics[width = 0.32\textwidth]{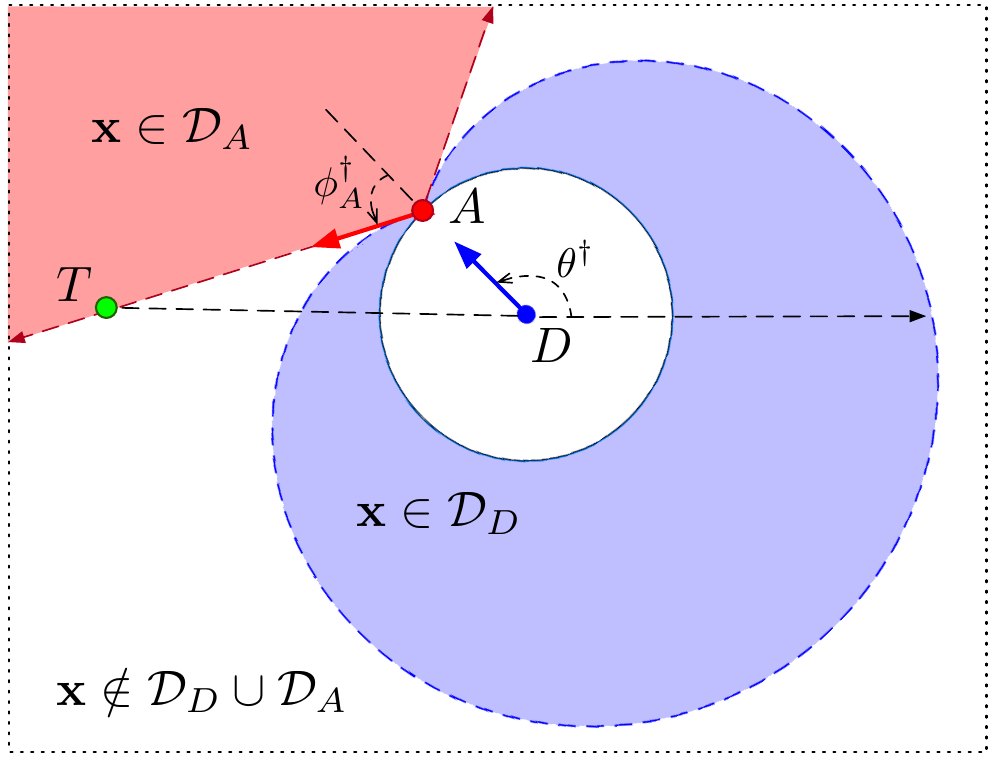}
    \caption{Attacker-win terminal condition: $\theta = \theta^\dagger(\rho_D)$ and the target position $\x{T} \in \mathcal{D}_A$.
    } 
    \label{fig: theta_crit}
\end{figure}

The attacker can win the game if it can safely `break free' from the defender's \edit{capture disk} and move towards the target in straight line
\edit{which is shown in Fig.~\ref{fig: theta_crit}.} 

\begin{lm}
     For any defender strategy, the attacker can leave the \edit{capture disk} and terminate Phase-II by employing $\phi_A\in(-\phi_A^\dagger, \phi_A^\dagger)$ where
\begin{equation}\label{eq: phi_A-crit}
    \phi_{A}^\dagger \triangleq \arccos{(\nu)}.
\end{equation} 
\end{lm}
\begin{proof}
    This result is easy to verify by considering the radial velocity of the attacker with respect to the defender:
    \edit{$\dot{R}
    =\cos{\phi_A}-\nu\cos{\phi_D}\geq\cos{\phi_A}-\nu$. }
The equality is achieved when the defender uses pure pursuit strategy ($\phi_D=0$). With $|\phi_A| < \phi_A^\dagger$, the radial velocity of the attacker is $\cos{\phi_A} >\nu$, which provides $\dot{R}>0$.
\end{proof}

\begin{rem}\label{rem: D_A}
When $R=r$, the attacker's heading tangent to CO is $\phi_A=\phi_A^\dagger$. The cone defined by $|\phi_A|\leq \phi_A^\dagger$ (see Figure~\ref{fig: theta_crit}) provides the attacker's dominance region $\mathcal{D}_A$.
\end{rem}
This result directly follows as a special case of Lemma~\ref{lemma: CO} and \eqref{eq: phi_S}.
The critical attacker position, $\theta^\dagger(\rho_D)$, that allows it to terminate Phase-II and reach the target is given in the following theorem.

\begin{theorem}[Attacker Win Condition]\label{thm: theta_dagger}
The attacker can terminate Phase-II and win the game if $|\theta| \geq \theta^\dagger$, where $\theta^\dagger$ is given as follows:
\edit{\begin{align}\label{eq: theta_crit}
    \theta^\dagger(\rho_D) \triangleq \pi -
    \arccos{(\nu)} + \arcsin{\left(\frac{r\sqrt{1-\nu^2}}{\rho_D}\right)}.
\end{align}
}
\end{theorem}
\begin{proof}
Recall that $\x{T}\in\mathcal{D}_A$ is a sufficient condition for the attacker to win. Now, based on Remark~\ref{rem: D_A}, it is clear that the attacker can win the game if the line of sight to the target is within the cone $|\phi_A|\leq\phi_A^\dagger$. The critical case is described in Figure~\ref{fig: theta_crit} where $\phi_A = \phi_A^\dagger$. The angle $\theta^\dagger$ can be described as $\theta^\dagger = \angle(TAD)+\angle(DTA)$, where $\angle(TAD)=\pi - \phi_A^\dagger$ and $\angle(DTA)$ can be found by the law of sine: 
\begin{align}
\edit{\frac{\sin{\angle(DTA)}}{r}  = \frac{\sin{\phi_A^\dagger}} {\rho_D} = \frac{\sqrt{1-\nu^2}}{\rho_D}.}
\end{align}
From \edit{Remark 3}, one can see that the attacker will immediately leave the \edit{capture disk}, and Phase-II terminates.
\end{proof}
The attacker's winning surface $\theta=\theta^\dagger$ is depicted by the red solid line in Figure~\ref{fig: terminal_surface}.

\subsection{Phase-II: Optimal Constrained Trajectories}
\label{sec:}

This section considers how the agents can steer the states $(\rho_D,\theta)$ towards the terminal surfaces corresponding to their respective win.

The strategy we consider for the attacker is to allocate sufficient radial velocity to stay on the \edit{capture disk}, according to \eqref{eq: p2_heading_constraint}, and use the remaining velocity component to circumnavigate (go around) the \edit{capture disk}.
For circumnavigation, the attacker intends to increase $|\theta|$, so that the target comes in its line of sight as discussed in Theorem~\ref{thm: theta_dagger}.
Although \eqref{eq: p2_heading_constraint} gives the attacker two options to circumnavigate in clockwise or counter-clockwise, only one results in increasing $|\theta|$.
This optimal attacker heading $\phi_A^*$  \edit{will} have the same sign as $\phi_D$, which means both the attacker and the defender moves either clockwise or counter-clockwise around the target.
Formally, we can write
\begin{equation}
    \phi_A^* = \text{sign}(\phi_D)\cdot\arccos(\nu\cos\phi_D).
\end{equation}
With this attacker strategy, $\dot{\theta}$ becomes a function of the defender's control input $\phi_D$:
\edit{\begin{equation}
\begin{aligned}
    \dot{\theta}(\phi_D) &= -\frac{\nu}{\rho_D}\sin{(\phi_D+\theta)} \\
    &+ \frac{1}{r}\left(\sqrt{1-\nu^2\cos^2{\phi_D}}-\nu\sin{\phi_D}\right). 
\end{aligned}
\end{equation}}

Recalling that $\dot{\rho}_D$ is solely controlled by the defender as in \eqref{eq: kinematics rhoDdot}, both $\dot{\theta}$ and $\dot{\rho_D}$ are now functions of $\phi_D$.
For the defender to decide how to steer the states in $(\rho_D,\theta)$ space, one can make the following observations:
\begin{itemize}
\item The state $(\rho_D,\theta_1)$ is more advantageous than $(\rho_D,\theta_2)$ for the defender if $|\theta_1|>|\theta_2|$. 
\item The state $(\rho_{D,1},\theta)$ is more advantageous than $(\rho_{D,2},\theta)$ for the defender if $\rho_{D,1}<\rho_{D,2}$. 
\end{itemize}
Based on these observations, we conclude that the defender wants to steer the state towards $\rho_D=r$ (approach the target) while maximizing $\theta$ (blocking the attacker).
To this end, the following objective function captures the slope of the trajectory in $(\rho_D,\theta)$ space:
\begin{align}\label{eq: objective_func}
     \frac{\dot{\theta}}{\dot{\rho}_D} 
    &= \frac{-\frac{\nu}{\rho_D}\sin{(\phi_D+\theta)} + \frac{1}{r}\left(\sqrt{1-\nu^2\cos^2{\phi_D}}-\nu\sin{\phi_D}\right)}{\nu\cos{(\phi_D+\theta)}},
\end{align}
and the optimal defender heading is given by the following equation:
\begin{equation}\label{eq: phi_D^*}
    \phi_D^* = \arg \min_{\phi_D} \left(\frac{\dot{\theta}}{\dot{\rho}_D}\right), 
\end{equation}
subject to $\dot{\rho}_D<0$. 
Since \eqref{eq: phi_D^*} cannot be solved for a closed-form expression of \(\phi_D^*\), \edit{we solve \(\phi_D^*\) numerically. It is important to choose the appropriate sign for \(\phi_D^*\) depending on whether the interaction between the defender and the attacker is clockwise or counter-clockwise.}



\begin{figure}
    \centering
    \includegraphics[width = 0.4\textwidth]{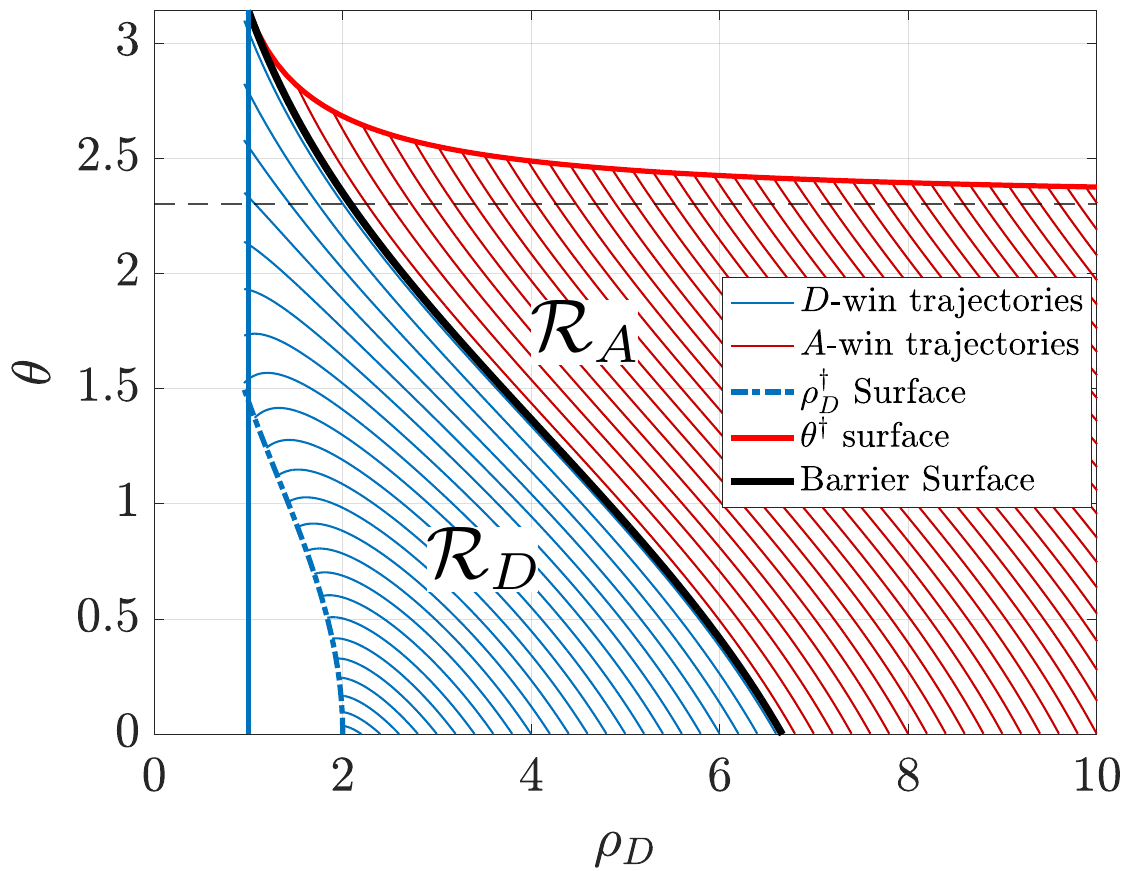}
    \caption{
    Phase-II trajectories in defender and attacker winning region, separated by the barrier surface indicated by the black line. 
    } 
    \label{fig: trajectories_p2}
\end{figure}
Given $\phi_D^*$ and $\phi_A^*$, Figure~\ref{fig: trajectories_p2} shows the trajectories in the $(\rho_D,\theta)$ space for Phase-II. 
The black line is the trajectory that will terminate at the intersection of defender-win and attacker-win surface: $(\rho_D,\theta)=(1,\pi)$. This trajectory serves as the barrier surface for phase-II.
This complete characterization of Phase-II serves as part of the terminal surface for Phase-I.

Although we are interested in solving the game of kind, for
convenience, we label these different trajectories with some
performance index as follows:
\begin{align}
    V(\rho_D,\theta) = \left\{
    \begin{array}{cc}
    \pi-\theta_f &\text{if\;} (\rho_D,\theta)\in \mathcal{R}_D\\
    -\rho_{D,f} & \text{otherwise.}
    \end{array}
    \right.
\end{align}
where $\rho_{D,f}$ and $\theta_f$ are the states at the final time of Phase-II. With this definition, $V(\rho_D,\theta)=0$ on the barrier, and $V>0$ (resp.~$V<0$) corresponds to $\mathcal{R}_D$ (resp.~$\mathcal{R}_A$). \edit{The choice of $V$ is motivated by the goal of ensuring that the defender (or attacker) not only aims to reach the target but also seeks to increase the distance between the target and the attacker (or defender).} Note that we do not intend to use this as a payoff. 

\subsection{Phase-I: Optimally Entering Phase-II}
\label{sec:asdfasdf}
\begin{figure}
    \centering
\includegraphics[width = 0.45\textwidth]{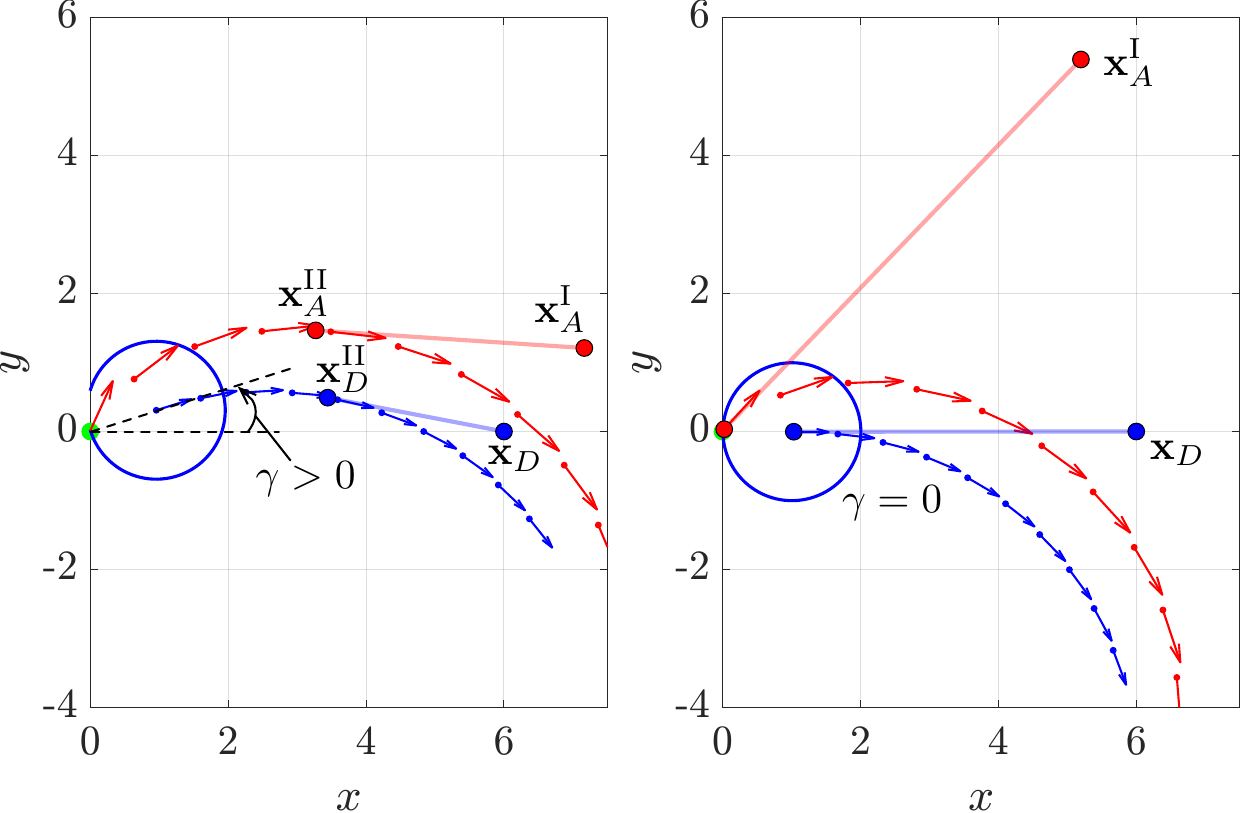}
    \caption{Computing attacker position on the barrier from retrograde Phase-II trajectories for $\gamma = 0.4\pi$ (left) and $\gamma = 0$ (right).
    }
    \label{fig: phase_trans}
\end{figure}

In this section, we derive the optimal trajectories of the players \edit{during} Phase-I. 
The derivation will rely on the smooth transition between Phase-I and Phase-II.  Although this ``property'' was \edit{initially utilized} in \cite{hagedorn1976differential} and also in \cite{fu2022defending}, we \edit{treat} it as a conjecture \edit{in the absence of a formal proof:}
\begin{conj}
The optimal trajectory in Phase-I smoothly connects to the optimal trajectory in Phase-II\edit{, ensuring that there is no discontinuity in the heading as the phase transition occurs.}
\end{conj} 

The trajectories in Phase-II can be \edit{represented} in the $xy$-plane to \edit{visualize} the physical \edit{paths} of the players. 
Due to the rotational symmetry in Phase-II analysis, we have a set of trajectories each corresponding to a particular terminal heading angle in $xy$-plane.
We use $\gamma$ to denote this rotational degree of freedom: which can be defined as the angle between $\x{T/D} \triangleq \x{T}-\x{D}$
and $\dot{\mathbf{x}}_D(t_1)$, where $t_1$ is the initial time of Phase-I.
See Figure~\ref{fig: phase_trans} for two trajectories both corresponding to $V=0$: one with $\gamma=0$ and the other with some $\gamma>0$.

The Phase-II trajectories in $xy$-plane can be parameterized using $V$ and $\gamma$.
For each Phase-II trajectory, we can identify the tangent line from the current defender position. 
Based on Conjecture~1, this gives us the location, $\x{D}^\text{II}(V,\gamma)$, that the defender enters Phase-II. It also gives us the duration of Phase-II, denoted as $\tau_\text{II}$, as well as the duration of Phase-I: which can be computed as $\tau_\text{I} \triangleq \|\x{D}^\text{II}(V,\gamma)-\x{D}\|/\nu$.

Using $\tau_\text{II}$, we can identify the corresponding location 
\edit{at which the attacker transitions into Phase-II,} denoted by $\x{A}^\text{II}(V,\gamma)$, 
\edit{along with the} velocity vector $\vvec{A}^\text{II}(V,\gamma)$ at that point.
The position of the attacker for this particular Phase-II trajectory to be valid is given by:
\begin{equation}
\label{eq: xA_V_gamma}
    \x{A}^\text{I}(V,\gamma;\x{D}) = \x{A}^\text{II} - \tau_\text{I}\cdot \vvec{A}^\text{II},
\end{equation}
where the dependency of each variables on $(V,\gamma)$ and $\x{D}$ is omitted on the right-hand side for clarity.

For an initial condition given by $(\x{D},\x{A}^\text{I}(V,\gamma;\x{D}))$, we know by construction that each agent's optimal strategy is to move in a straight line towards the tangent point of Phase-II trajectory with the value of $V$ and the rotation of $\gamma$.
Based on this result, one can numerically search over $(V,\gamma)$ space to identify the one that matches the actual attacker position.


\if{false}
\subsubsection{Barrier}
For the barrier, the trajectories of Phase-I can be found by the tangent lines connecting the defender's initial position $\x{D,0}$ and the Phase-II trajectories \cite{fu2022defending},\cite{hagedorn1976differential}. 

The barrier surface of the blocking game in Phase-I is a set of points from which the attacker goes through Phase-II to reach the target critically. 
The trajectories of the players can be found by combining Phase-I and Phase-II trajectories which are obtained in retrograde manner. In Phase-I the trajectories are straight line connecting Phase-II trajectory tangentially which is shown in Figure~\ref{fig: phase_trans}.

Let the time it takes for the defender to reach the Phase-II trajectory at $\x{D,II}$ from the initial position $\x{D,0}$ be denoted as $t_I$. 

The trajectories in Phase-II can be plotted in $xy$-plane to get the physical trajectories of the players. Note that the trajectories in the $(\rho_D,\theta)$ space has the rotational symmetry around the target. Therefore we can get different Phase-II trajectories in $xy$-plane. In polar coordinate, let $(r,\gamma)$ represents defenders position in terminal time of the Phase-II, where $\gamma$ is the angular position of the defender in $xy$-plane, shown in Figure~\ref{fig: phase_trans}. For a given $\gamma$, the Phase-II trajectories are found from $(\rho_D,\theta)$ trajectories. 

For a given defender initial position $\x{D,0}$ in Phase-I, let defender reach in Phase-II at $\x{D,II}^\gamma$ and corresponding attacker position $\x{A,II}^\gamma$, where $\|\x{D,II}^\gamma-\x{A,II}^\gamma\| = r$. Let $t_I$ be the time players are in Phase-I such that $R > r, \mkern3mu \forall t \in (0,t_I)$ where
\begin{align}
    t_I = \|\x{D,II}^\gamma-\x{D,0}\|/\nu.
\end{align}
We can compute the attacker's initial position in Phase-I, $\x{A,0}^\gamma$ as follows:
\begin{align}\label{eq: A_II,0_gamma}
\x{A,0}^\gamma = \x{A,II}^\gamma - t_I \cdot v_A,
\end{align}
where $v_A \in [\cos{\phi}_A, \sin{\phi}_A]^\top$ and $\phi_A$ 
is the attacker heading tangentially connecting Phase-II trajectory at $\x{A,II}^\gamma$ to $\x{A,0}^\gamma$. 
\fi

\section{Barrier and Security Strategies}
We combine the results of Sec.~\ref{sec: dominance} and Sec.~\ref{sec: IV blocking game} to construct the full barrier surface in $\mathbf{z}$, and provide a security strategy for each agent 
\edit{within their} corresponding winning \edit{regions}. 

\subsection{Barrier Surface}

We first define the \emph{natural barrier}, \edit{derived} from the \edit{results in} Sec.~\ref{sec: dominance}.
When blocking does not happen, the game does not enter Phase-II. 
This is true if $\angle ATD>\arccos{(\nu)}$~\cite{fu2022defending}.
In this case, the attacker is able to win if the following condition is satisfied: $\|\x{A}\| < \|\x{D}\|/\nu$. 
The critical case provides the natural barrier as follows:
\begin{align}
    \mathcal{B}_N = \{\x{A} \mid \|\x{A}\| = \|\x{D}\|/\nu \;\wedge\; \angle ATD \geq \arccos(\nu) \}.
\end{align}
\begin{rem}\label{rem: Bbar}
    The condition $\|\x{A}\| = \|\x{D}\|/\nu$ is sufficient for the defender to win if $\angle ATD < \arccos(\nu)$. This is because the attacker must now circumnavigate the defender, even if the defender is moving straight to the target. The attacker will spend more time than $\|\x{D}\|/\nu$ to reach the target. We denote this part of the circle as $\bar{\mathcal{B}}_N$, later used in
    Theorem~\ref{thm: Dwin}.
\end{rem}

From \eqref{eq: xA_V_gamma}, we can identify the set of attacker positions on the barrier by fixing $V=0$ and varying $\gamma$.
One can verify that $\gamma=0$ corresponds to the critical point where the natural barrier ends. As $\gamma$ is increased, the attacker position moves, and it either hits the \edit{capture disk} as in Figure~\ref{fig: phase_trans} \edit{(left)}, or hits the line of symmetry: co-linear with target and defender as in Figure~\ref{fig: phase_trans} \edit{(right)}. Denoting this angle as $\gamma_\text{max}$, we can define the counter-clockwise part of the \emph{envelope barrier} as:
\begin{align}
    \mathcal{B}_E =\{\x{A} \mid \x{A}^\text{I}(0,\gamma,\x{D}),\gamma\in[0,\gamma_\text{max}]\}. 
\end{align}
The clockwise part can be obtained similarly.

\begin{figure}
    \centering
    \includegraphics[width = 0.49\textwidth]{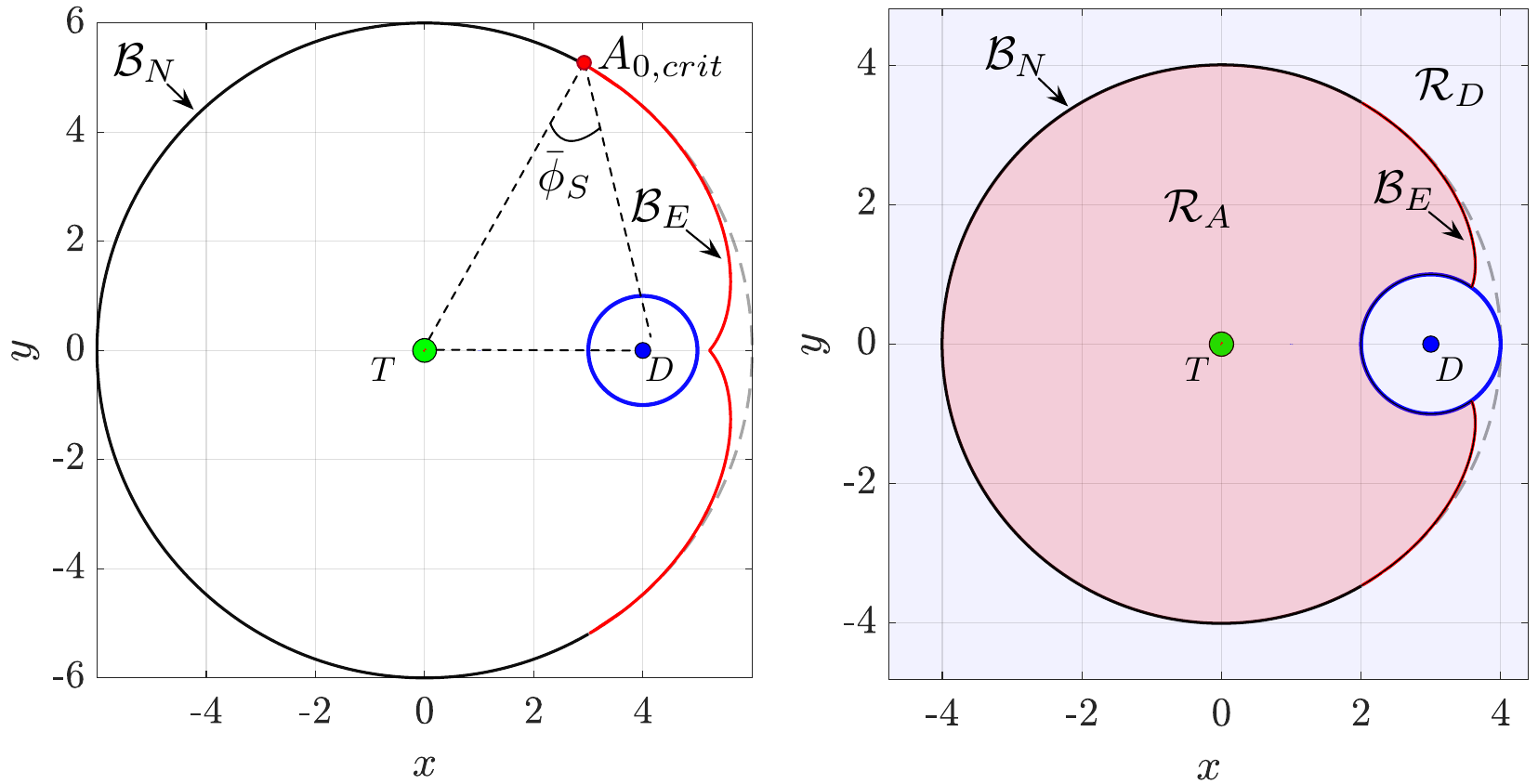}
    \caption{Winning regions of the attacker and defender, separated by the natural barrier, $\mathcal{B}_N$ and the envelope barrier, $\mathcal{B}_E$ \edit{for $\nu = 0.5, r=1$, $\x{D} = [4,0]^\top$ (left), and $\x{D} = [3,0]^\top$ (right). }} 
    \label{fig: construction_of_Barrier}
\end{figure}

The overall barrier is attained by the combination of $\mathcal{B}_N$ and $\mathcal{B}_E$ as shown in Figure~\ref{fig: construction_of_Barrier}. Since the two surfaces smoothly connect with each other, they always provide a closed region, which is the attacker winning region $\mathcal{R}_A$, and the exterior is the defender winning region $\mathcal{R}_D$.

\if false
From \eqref{eq: xA_V_gamma}, we can identify the set of attacker positions on the barrier by fixing $V=0$ and varying $\gamma$:
\begin{align}
    \mathcal{B}_e =\{\x{} \mid \|\x{A}-\x{A,II}^\gamma\| = \|\x{A,0}-\x{A,II}^\gamma\| \}. 
\end{align}
At the critical case $\gamma = \gamma^\dagger$, when 
$x_{A,II}^{\gamma} = [0,0]^\top$ the corresponding $\x{A,0}^\gamma$ dictates that the target is reachable by following a straight-line trajectory to the target. 
For $\gamma < \gamma^\dagger$, the points obtained by connecting Phase-I and Phase-II trajectories form the envelope barrier, $\mathcal{B}_e$. 

When $\gamma \geq \gamma^\dagger$ the attacker does not go through Phase-II, and corresponding attacker position form the `natural barrier', $\mathcal{B}_n$: 
\begin{align}
    \mathcal{B}_n = \{\x{} \mid \|\x{A}\| = \|\x{D,0}\|/\nu \}
\end{align}
The transition between these two scenarios is shown in Figure~\ref{fig: construction_of_Barrier}. 

The attacker positions on the envelope barrier are shown by the red curve and the natural barrier which forms a circle are also shown by the black circle. The barriers for the clockwise and counterclockwise trajectories of Phase-II game, are connected.

The overall barrier can be given as follows:
\begin{align}
    \mathcal{B} = \begin{cases}
        &\mathcal{B}_e \quad\text{ when } \gamma <\gamma^\dagger, \\
        &\mathcal{B}_n  \quad\text{ when } \gamma \geq \gamma^\dagger.
    \end{cases}
\end{align}
\fi

\subsection{Security Strategies}
We summarize the main result by providing a security strategy for each player to win the game in its corresponding winning region.
\begin{theorem}[Defender win scenario]\label{thm: Dwin}
    If $\x{A}\in$ $\mathcal{R}_D$, the defender wins the game against any admissible attacker strategy by using the following strategy: 
    \begin{itemize}
    \item Move straight towards the target if $\|\x{A}\| \geq \|\x{D}\|/\nu$,

    \item Use the heading given by $\phi_D^*$ defined in \eqref{eq: phi_D^*} if $R=r$,

    \item Move straight towards $\x{D}^\text{II}(V,\gamma)$ for some $V>0$ and $\gamma$ that achieves $\x{A}^{I}(V,\gamma;\x{D})=\x{A}$, if $\|\x{A}\| < \|\x{D}\|/\nu$ and $R>r$.
    
    \end{itemize}
\end{theorem}
\begin{proof}
From the construction of the natural barrier and Remark~\ref{rem: Bbar}, the case with $\|\x{A}\| \geq \|\x{D}\|/\nu$ is trivial.
Based on the Phase-II analysis, and the corresponding optimal strategy, the case with $R=r$ is also straightforward.
The remaining question is what happens in the area outside of the envelope barrier, but inside the circle defined by $\|\x{A}\| = \|\x{D}\|/\nu$.

Suppose that the attacker position corresponds to $V=V_0>0$ and some $\gamma$ at the initial time.
The defender can move straight towards $\x{D}^\text{II}(V_0,\gamma)$ to ensure that the attacker does not penetrate the level set  defined by
\begin{align}
    \mathcal{B}_L(V_0) =\{\x{A} \mid \x{A}^\text{I}(V_0,\gamma,\x{D}),\gamma\in[0,\gamma_\text{max}]\}. 
\end{align}
In this case, one of the following two will happen: (i) the game reaches Phase-II with states $(\rho_D,\theta)$ in defender-winning region corresponding to $V\geq V_0>0$; or (ii) the attacker touches $\bar{\mathcal{B}}_N$, which ensures defender's win from Remark~\ref{rem: Bbar}.
\end{proof}

The attacker's winning strategy can be constructed similarly.
\begin{theorem}[Attacker win scenario]
    If $\x{A}\in \mathcal{R}_A$, the attacker wins the game against any admissible defender strategy by following the following strategy:
    \begin{itemize}
    \item Move straight towards the target if $\x{T}\in\mathcal{D}_A$,
    
    \item Use the distance maintaining strategy according to $\phi_A^*$ if $\x{T}\notin\mathcal{D}_A$ and $R=r$,

    \item Move straight towards $\x{A}^\text{II}(V,\gamma;\x{D})$ for some $V<0$ and $\gamma$ that achieves $\x{A}^{I}(V,\gamma;\x{D})=\x{A}$, if $\x{T}\notin\mathcal{D}_A$ and $R>r$.
    \end{itemize}
\end{theorem}
\begin{proof}
We have already proved the case with $\x{T}\in\mathcal{D}_A$ in Lemma~\ref{lem: D_A}. 
Based on the Phase-II analysis and the corresponding optimal strategy, the case with $R=r$ is also straightforward.
In the third case, we can proceed with a similar argument as in Theorem~\ref{thm: Dwin}.
The attacker can ensure that it does not cross the level set $\mathcal{B}_L(V_0)$ for the initial $V_0<0$.
This implies that the attacker either: (i) reaches Phase-II with states $(\rho_D,\theta)$ corresponding to attacker winning region; or (ii) attains the condition $\x{T}\in\mathcal{D}_A$.
\end{proof}


\subsection{Singular Surface} 
For SD-TDG there exists one singular surface on which the optimal strategy discussed in the paper is nonunique. 
This occurs when the target, defender, and the attacker are in a co-linear formation i.e., $\theta = 0$. In this case the defender can either choose to move clockwise or counterclockwise to potentially block the attacker. 
Likewise, the attacker can also choose from two control actions that are solutions derived from the game. In both cases if the attacker chooses the opposite of what defender chooses, the defender incurs a small loss.

Noting that the co-linear configuration (i.e., $\theta=0$) is unfavorable for the attacker (since it must perform full circumnavigation to reach the target), the attacker will not attempt to steer the system back to this singular surface.
On the other hand, the defender does not have the control authority to steer $\theta$, unless the condition \eqref{eq: D_perp block} is satisfied.
Therefore, the game will not encounter perpetuated dilemma.
The formal proof of this claim is out of the scope of this conference paper.

\section{Simulation}

\begin{figure}
    \centering
     \includegraphics[width =0.32\textwidth]{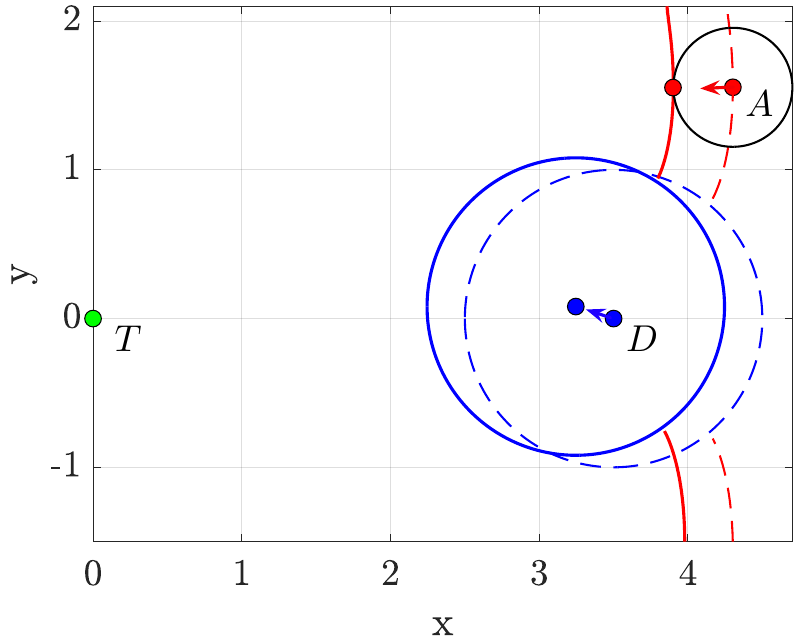}
    \caption{\edit{Visualization of equilibrium headings and barrier surface in Phase-I for }
    \edit{$\x{D}(0) = [3.5,0]^\top,
    \x{A}(0) = [  4.3021, 1.5550]^\top$.}
    } 
    \label{fig: optimality_check_in_RD}
\end{figure}

\begin{figure}
    \centering
    \includegraphics[width = 0.38\textwidth]{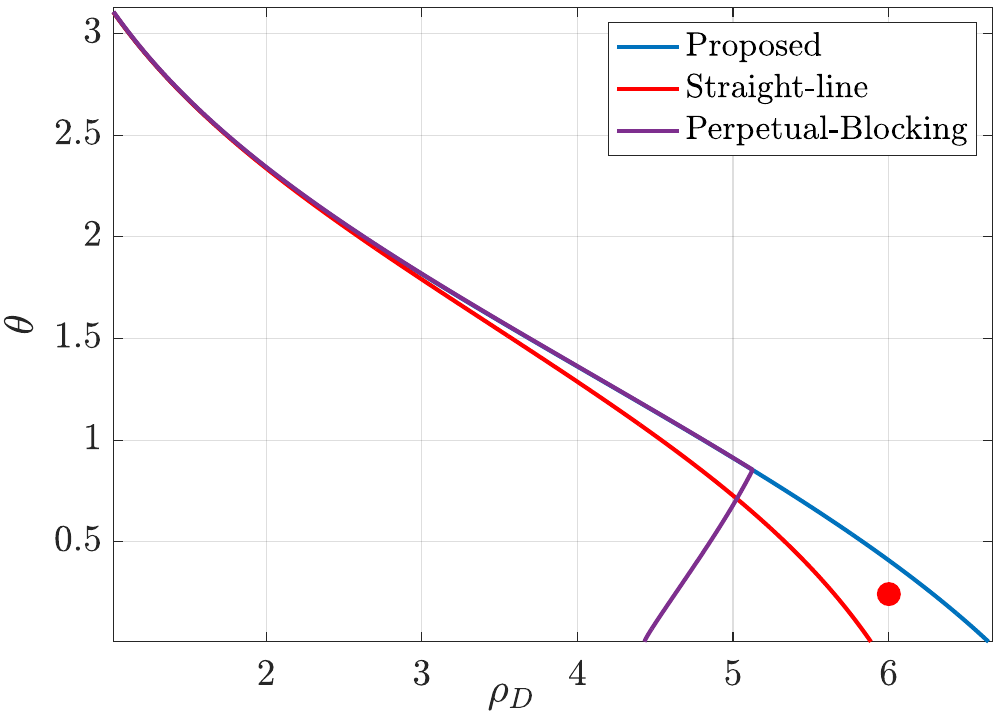}
    \caption{Comparison between the Phase-II barrier surface for
    open-loop straight-line strategy \cite{pachter2022strategies}, perpetual blocking\cite{fu2021optimal}, and  the proposed strategy in $(\rho_D,\theta)$ space. 
    } 
    \label{fig: comparison}
\end{figure}
\begin{figure}
    \centering
    \includegraphics[width = 0.4\textwidth]{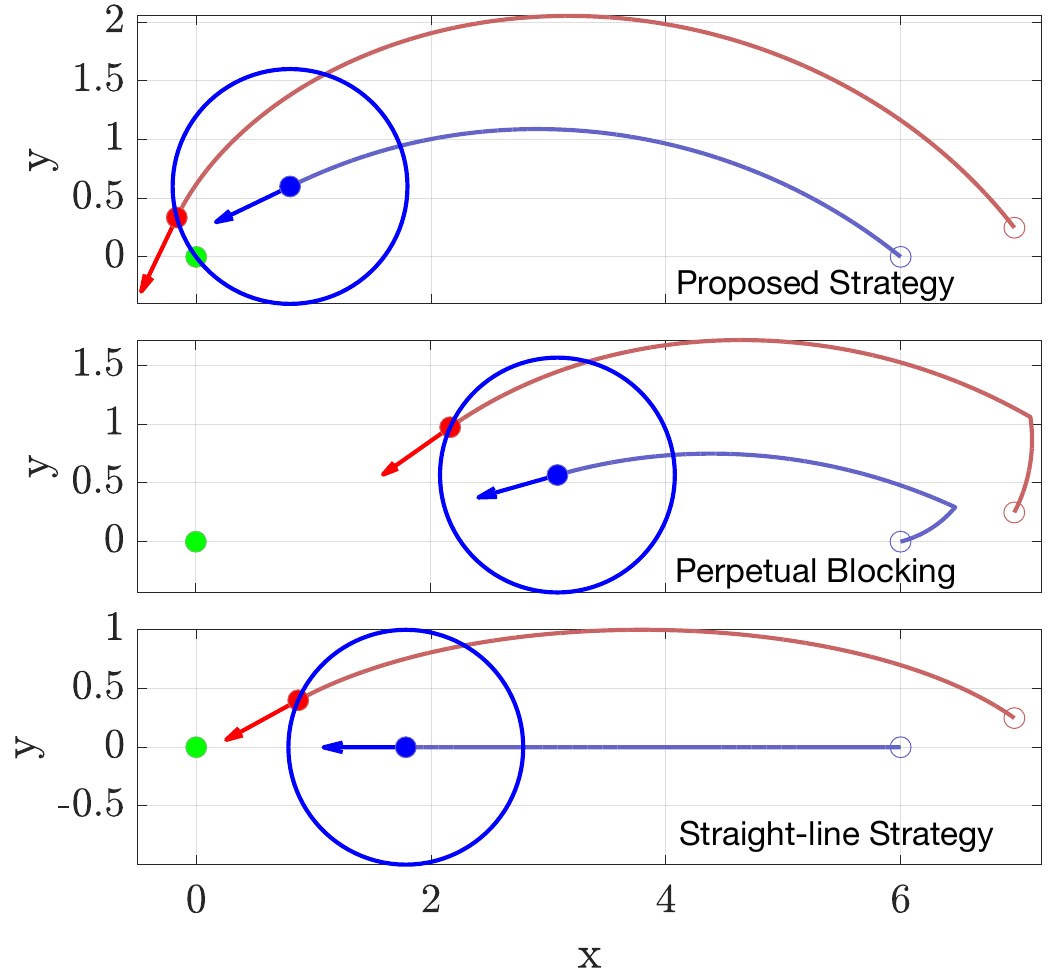}
    \caption{Comparison between the trajectories of the $A$ and $D$ for different defender strategies.
    } 
    \label{fig: comparison_hf_vs_proposed}
\end{figure}
In this section we first verify the optimaility of the proposed strategies in Phase-I. We choose following initial conditions and parameters: 
    $\x{D}(0) = [3.5,0]^\top, \nu = 0.75$. We choose an \edit{arbitrary} attacker \edit{position} on the barrier at $\x{A}(0) = [4.30, 1.56]^\top$.
    We obtain the optimal controls for the players as follows: $\vvec{D} = [-0.95,0.30]^\top,  \vvec{A} = [-1.0,0]^{\top}$.
    Figure~\ref{fig: optimality_check_in_RD} shows the barrier after $\Delta t=0.4$ as well as all possible attacker positions after this duration (indicated as a black disk). 
    It shows that the attacker can, at best, stay on the barrier following the optimal strategy, but it cannot penetrate the barrier; verifying the optimality of defender's Phase-I strategy.

Now we discuss three different defender strategies in Phase-II of the game. 
We consider an open-loop ``straight-line'' strategy suggested in \cite{pachter2022strategies}, ``perpetual-blocking" strategy suggested in \cite{fu2022defending,fu2021optimal} and the proposed strategy in this paper. Figure~\ref{fig: comparison} shows the barrier of the Phase-II game following these strategies. 
Notice that the region below each curve corresponds to the defender winning condition.
The figure shows that this defender-win region is the largest with our proposed strategy, and it contains the ones from other strategies.
The red dot in Figure~\ref{fig: comparison} gives an example of the initial condition that will lead to defender's win with our strategy, but not with the other two.
In $xy$-plane this initial condition corresponds to $\x{D} = [6,0]^\top$, $\x{A} = [6.97,0.25]^\top$, 
$\nu = 0.75$, and $r = 1$. 

The trajectories of the players from the above initial condition are shown in Figure~\ref{fig: comparison_hf_vs_proposed}.
For the open-loop straight-line strategy\cite{pachter2022strategies}, the attacker circumnavigates around the defender and finally have a line-of-sight to the target that leads the attacker win.  
Using the strategy described in \cite{fu2022defending}, \cite{fu2021optimal} the defender initially pushed the attacker farther from the target, however the attacker loops-around the defender and finally have the line-of sight to the target, thus wins. 
Finally, following our proposed strategy, the defender blocks the attacker until it reaches the target and wins the game.

\edit{These results demonstrate that our proposed defender strategy is superior to the other two.}
The improvement is achieved because our strategy is tailored for a point-target scenario, whereas the one in \cite{fu2022defending}, \cite{fu2021optimal} is more appropriate for target regions that can only be guarded by perpetually circling around it.

\section{Conclusion}

In this paper we studied target defense game using a slower defender. 
We first use dominance regions to carve out the portion of the state space resulting in trivial winning conditions.
A more interesting scenario where the defender can ``block'' the attacker is studied by decomposing the problem into three phases.
The optimal strategies on a constrained surface is solved first, and then the optimal way to approach that surface is derived.
We provide the solution to the game of kind by constructing the barrier surface and providing a security strategy for each player to win in its corresponding winning region.
In future we would like to explore the game of degree which provides optimal strategies  within the winning regions. 
\typeout{} 
\bibliographystyle{IEEEtran}
\bibliography{citation}

\begin{thebibliography}{10}
\providecommand{\url}[1]{#1}
\csname url@samestyle\endcsname
\providecommand{\newblock}{\relax}
\providecommand{\bibinfo}[2]{#2}
\providecommand{\BIBentrySTDinterwordspacing}{\spaceskip=0pt\relax}
\providecommand{\BIBentryALTinterwordstretchfactor}{4}
\providecommand{\BIBentryALTinterwordspacing}{\spaceskip=\fontdimen2\font plus
\BIBentryALTinterwordstretchfactor\fontdimen3\font minus
  \fontdimen4\font\relax}
\providecommand{\BIBforeignlanguage}[2]{{%
\expandafter\ifx\csname l@#1\endcsname\relax
\typeout{** WARNING: IEEEtran.bst: No hyphenation pattern has been}%
\typeout{** loaded for the language `#1'. Using the pattern for}%
\typeout{** the default language instead.}%
\else
\language=\csname l@#1\endcsname
\fi
#2}}
\providecommand{\BIBdecl}{\relax}
\BIBdecl

\bibitem{von2021turret}
A.~Von~Moll, D.~Shishika, Z.~Fuchs, and M.~Dorothy, ``The
  turret-runner-penetrator differential game,'' in \emph{2021 American Control
  Conf. (ACC)}.\hskip 1em plus 0.5em minus 0.4em\relax IEEE, 2021, pp.
  3202--3209.

\bibitem{weintraub2020introduction}
I.~E. Weintraub, M.~Pachter, and E.~Garcia, ``An introduction to
  pursuit-evasion differential games,'' in \emph{2020 American Control Conf.
  (ACC)}.\hskip 1em plus 0.5em minus 0.4em\relax IEEE, 2020, pp. 1049--1066.

\bibitem{pachter2020capture}
M.~Pachter, D.~W. Casbeer, and E.~Garcia, ``Capture-the-flag: A differential
  game,'' in \emph{2020 IEEE Conference on Control Technology and Applications
  (CCTA)}.\hskip 1em plus 0.5em minus 0.4em\relax IEEE, 2020, pp. 606--610.

\bibitem{Issacs1965}
R.~Isaacs, \emph{Differential Games: A Mathematical Theory with Applications to
  Optimization, Control and Warfare.}\hskip 1em plus 0.5em minus 0.4em\relax
  Wiley, New York, 1965.

\bibitem{shishika2021partial}
D.~Shishika, D.~Maity, and M.~Dorothy, ``Partial info. target defense game,''
  in \emph{2021 IEEE Inter. Conf. on Robotics and Automation (ICRA)}.\hskip 1em
  plus 0.5em minus 0.4em\relax IEEE, 2021, pp. 8111--8117.

\bibitem{pachter2019toward}
M.~Pachter, E.~Garcia, and D.~W. Casbeer, ``Toward a solution of the active
  target defense differential game,'' \emph{Dynamic Games and Applications},
  vol.~9, pp. 165--216, 2019.

\bibitem{VonMoll2020BD}
A.~Von~Moll, E.~Garcia, D.~Casbeer, M.~Suresh, and S.~C. Swar,
  ``Multiple-pursuer, single-evader border defense differential game,''
  \emph{Journ. of Aerospace Info. Syst.}, vol.~17, no.~8, pp. 407--416, 2020.

\bibitem{das2022guarding}
G.~Das and D.~Shishika, ``Guarding a translating line with an attached
  defender,'' in \emph{2022 American Control Conference (ACC)}.\hskip 1em plus
  0.5em minus 0.4em\relax IEEE, 2022, pp. 4436--4442.

\bibitem{shishika2019perimeter}
D.~Shishika and V.~Kumar, ``Perimeter-defense game on arbitrary convex
  shapes,'' \emph{arXiv preprint arXiv:1909.03989}, 2019.

\bibitem{lee2021guarding}
Y.~Lee and E.~Bakolas, ``Guarding a convex target set from an attacker in
  euclidean spaces,'' \emph{IEEE Control Syst. Lett.}, vol.~6, pp. 1706--1711,
  2021.

\bibitem{pourghorban2023target}
A.~Pourghorban and D.~Maity, ``Target defense against periodically arriving
  intruders,'' \emph{arXiv preprint arXiv:2303.05577}, 2023.

\bibitem{Garcia2021Multiple}
E.~Garcia, D.~W. Casbeer, A.~Von~Moll, and M.~Pachter, ``Multiple pursuer
  multiple evader differential games,'' \emph{IEEE Trans. on Automatic
  Control}, vol.~66, no.~5, pp. 2345--2350, 2021.

\bibitem{garcia2021cooperative_containment}
E.~Garcia and S.~D. Bopardikar, ``Cooperative containment of a high-speed
  evader,'' in \emph{2021 American control Conf. (ACC)}.\hskip 1em plus 0.5em
  minus 0.4em\relax IEEE, 2021, pp. 4698--4703.

\bibitem{ramana2015cooperative}
M.~Ramana and M.~Kothari, ``A cooperative pursuit-evasion game of a high speed
  evader,'' in \emph{2015 54th IEEE Conf. on Decision and Control (CDC)}.\hskip
  1em plus 0.5em minus 0.4em\relax IEEE, 2015, pp. 2969--2974.

\bibitem{vechalapu2020trapping}
T.~R. Vechalapu, ``A trapping pursuit strategy for capturing a high speed
  evader,'' in \emph{AIAA Scitech 2020 forum}, 2020, p. 2069.

\bibitem{fang2020cooperative}
X.~Fang, C.~Wang, L.~Xie, and J.~Chen, ``Cooperative pursuit with multi-pursuer
  and one faster free-moving evader,'' \emph{IEEE Trans. on Cyber.}, vol.~52,
  no.~3, pp. 1405--1414, 2020.

\bibitem{garcia2021cooperative_targetprotection}
E.~Garcia, ``Cooperative target protection from a superior attacker,''
  \emph{Automatica}, vol. 131, p. 109696, 2021.

\bibitem{pachter2022strategies}
M.~Pachter, D.~W. Casbeer, and E.~Garcia, ``Strategies for target defense from
  a fast attacker,'' in \emph{2022 IEEE Conf. on Control Tech. and Applications
  (CCTA)}.\hskip 1em plus 0.5em minus 0.4em\relax IEEE, 2022, pp. 233--238.

\bibitem{fu2022defending}
H.~Fu and H.~H.-T. Liu, ``Defending a target area with a slower defender,''
  \emph{IEEE Control Syst. Lett.}, vol.~7, pp. 661--666, 2022.

\bibitem{fu2021optimal}
------, ``Optimal solution of a target defense game with two defenders and a
  faster intrude,'' \emph{Unmanned Syst.}, vol.~9, no.~03, pp. 247--262, 2021.

\bibitem{hagedorn1976differential}
P.~Hagedorn and J.~Breakwell, ``A differential game with two pursuers and one
  evader,'' \emph{Journ. of Optimization Theory and Applications}, vol.~18, pp.
  15--29, 1976.

\end{thebibliography}
\end{document}